\documentclass[12pt,letterpaper]{article}
\usepackage{amsfonts}
\usepackage{amssymb}
\usepackage{amsmath}
\usepackage{amsthm}
\usepackage{multirow}
\usepackage[round,authoryear]{natbib}
\usepackage[margin=1in]{geometry}
\usepackage{booktabs}
\usepackage{dcolumn}
\newcolumntype{d}[1]{D{.}{.}{#1}} 
\usepackage{graphicx,epstopdf}
\usepackage[T1]{fontenc}
\usepackage{lmodern}
\usepackage{enumitem}
\setlist[enumerate,1]{itemsep=1pt, topsep=4pt, partopsep=0pt}
\setlist[enumerate,2]{nosep}
\setlist[itemize,1]{itemsep=1pt, topsep=4pt, partopsep=0pt}
\setlist[itemize,2]{nosep}
\usepackage{subfig}
\usepackage{bm}
\usepackage[colorlinks,linkcolor=red,citecolor=blue,pdftex,breaklinks]{hyperref}
\hypersetup{bookmarksopen=true}
\usepackage[nameinlink]{cleveref}
\usepackage{setspace}

\makeatletter
\AddToHook{env/figure/begin}{\def\ltx@label#1{\cref@label{#1}}}
\AddToHook{env/table/begin}{\def\ltx@label#1{\cref@label{#1}}}
\makeatother

\theoremstyle{plain}

\theoremstyle{definition}

\newtheorem{examplex}{Example}
\newenvironment{example}{\pushQED{\qed}\examplex}{\popQED\endexamplex}

\Crefname{assumptionx}{Assumption}{Assumptions} 
\Crefname{examplex}{Example}{Examples} 
\Crefname{remarkx}{Remark}{Remarks} 

\makeatletter
\renewcommand*{\eqref}[1]{\hyperref[{#1}]{\textup{\tagform@{\ref*{#1}}}}}
\makeatother

\makeatletter
\DeclareRobustCommand\citepos
{\begingroup\def\NAT@nmfmt##1{{\NAT@up##1's}}%
\NAT@swafalse\let\NAT@ctype\z@\NAT@partrue
\@ifstar{\NAT@fulltrue\NAT@citetp}{\NAT@fullfalse\NAT@citetp}}
\makeatother

\expandafter
\def \expandafter \normalsize \expandafter{\normalsize \setlength \abovedisplayskip{10pt plus 2pt minus 7pt}}
\expandafter
\def \expandafter \normalsize \expandafter{\normalsize \setlength \abovedisplayshortskip{0pt plus 2pt}}
\expandafter
\def \expandafter \normalsize \expandafter{\normalsize \setlength \belowdisplayskip{10pt plus 2pt minus 7pt}}
\expandafter
\def \expandafter \normalsize \expandafter{\normalsize \setlength \belowdisplayshortskip{5pt plus 2pt minus 3pt}}

\def\biW{{\bm{W}}}

\def\biS{{\bm{S}}}

\def\bSigma{{\bm{\Sigma}}}

\def\bPsi{{\bm{\Psi}}}

\def\E{\mathbb{E}}

\def\IF{{\mathbb I}}

\def\N{{\rm N}}
\def\dto{\overset{d}  \longrightarrow}

\def\tk{\kern 0.08333em}
\def\tn{\kern -0.08333em}
\def\tkk{\kern 0.04167em}
\def\bzero{{\bm{0}}}

\def\ATT{\mbox{ATT}}

\newcommand{\FO}[1]{}

\begin{document}

\def\rnd{\textcircled{r}}

\title{Improved Inference for CSDID\\Using the Cluster
Jackknife\thanks{We are grateful to audience members at the 2025
Canadian Stata Conference for helpful suggestions. MacKinnon and Webb
thank the Social Sciences and Humanities Research Council of Canada
(SSHRC grants 435-2025-1150 and 435-2021-0396) for financial support.
MacKinnon and Nielsen thank the Aarhus Center for Econometrics (ACE)
funded by the Danish National Research Foundation grant number DNRF186
for financial support. Webb thanks the Canadian Institute For Health
Research for financial support. Karim thanks the Ontario Graduate
Scholarship for financial support. Thanks to Yunhan (Max) Liu for
outstanding research assistance.\par
\rnd\ Certified random order: RzcE5N1cRUMb.}}

\author{Sunny R. Karim$^{\rm a}$ \rnd\\ \texttt{SunnyKarim@cmail.carleton.ca}
\and Morten \O rregaard Nielsen$^{\rm b}$ \rnd\\ \texttt{mon@econ.au.dk} 
\and James G. MacKinnon$^{{\rm b,c}}$ \rnd\\ \texttt{mackinno@queensu.ca} 
\and Matthew D. Webb$^{\rm a,}$\thanks{Corresponding
author.}\hspace*{10pt}\rnd\\
\texttt{matt.webb@carleton.ca}}
	
\date{%
$^{\rm a}$Department of Economics, Carleton University\\%
$^{\rm b}$Aarhus Center for Econometrics, Aarhus University\\%
$^{\rm c}$Department of Economics, Queen's University\\[2ex]%
\today
}
	
\maketitle

\begin{abstract}
Obtaining reliable inferences with traditional
difference-in-differences (DiD) methods can be difficult. Problems can
arise when both outcomes and errors are serially correlated, when
there are few clusters or few treated clusters, when cluster sizes
vary greatly, and in various other cases. In recent years, recognition
of the ``staggered adoption'' problem has shifted the focus away from
inference towards consistent estimation of treatment effects. One of
the most popular new estimators is the CSDID procedure of
\citet{callaway2021difference}. We find that the issues of
over-rejection with few clusters and/or few treated clusters are at
least as severe for CSDID as for traditional DiD methods. We also
propose using a cluster jackknife for inference with CSDID, which
simulations suggest greatly improves inference. We provide software
packages in \texttt{Stata} (\texttt{csdidjack}) and \texttt{R}
(\texttt{didjack}) to calculate cluster-jackknife standard errors
easily.
\end{abstract}

\clearpage

\section{Introduction}

The difference-in-differences (or DiD) model has been a workhorse of
empirical economics for several decades. Until about five years ago,
treatment effects for DiD models were typically estimated using
two-way fixed effects (TWFE) regressions. Recently, a series of highly
influential papers
\citep*{de2020twott,callaway2021difference,goodman2021difference} has
shown that TWFE regressions only recover the intended treatment
effects when some very strong assumptions hold. Several of these
assumptions fail to hold in a great many empirical applications. TWFE
regressions often yield inconsistent estimates of the desired
treatment effects because they involve ``forbidden comparisons'' and
``negative weights.'' New methods have therefore been proposed, and
thousands of papers have been written using them. The focus of this
new literature has been on estimating treatment effects correctly, but
relatively little attention has been paid to inference. Notable 
exceptions are \citet{weiss2024much} and
\cite{mizushima2025inference}. Perhaps the most widely-used of these
new methods is the one proposed by \citet{callaway2021difference},
which is often referred to as~CSDID.

It has been known for many years that inference in TWFE regressions
can be seriously unreliable when the disturbances are correlated
within clusters. \citet*{BDM_2004} popularized this issue, identified
serial correlation as one source of the problem, and suggested using
cluster-robust standard errors to improve inferences.
\citet*{CGM_2008} then showed that these standard errors can be
unreliable when the number of clusters in a sample is not large and
proposed the wild cluster bootstrap. The literature on cluster-robust
inference for conventional DiD models and other regression models has
exploded since then. \citet*{MNW-guide} is a recent guide.

Recently, it has been recognized that the cluster jackknife often
provides a convenient and reliable way to obtain cluster-robust
standard errors \citep*{MNW-bootknife,Hansen-jack,Hansen_2025}. The
first of these papers also introduces a jackknife-inspired
transformation to the wild cluster bootstrap, which can greatly
improve inferences.

Given the difficulty of making reliable cluster-robust inferences with
traditional TWFE regressions, it seems very likely that modern DiD
methods, including the CSDID estimator, will also encounter
difficulties. In \Cref{sec:MCR}, we show that they do. To improve
inference in finite samples, we therefore propose using the cluster
jackknife. The jackknife substantially improves the reliability of
inferences, but the problem of few treated clusters remains in some
settings. To facilitate the estimation of jackknife standard errors,
we provide both \texttt{Stata} and \texttt{R} packages. 

The remainder of the paper is organized as follows. The first part of
\Cref{sec:notation} introduces the notation, while
\Cref{sec:twfe} reviews the problem of staggered adoption for
TWFE models. \Cref{sec:cs} briefly reviews the CSDID estimator and the
associated inference procedures. \Cref{sec:cluster} describes the two
existing methods for cluster-robust inference with CSDID, and
\Cref{sec:jack} proposes the cluster jackknife. \Cref{sec:MCD}
describes the Monte Carlo design to compare the various procedures,
and \Cref{sec:MCR} discusses the results. \Cref{sec:examples} briefly
examines two empirical examples. \Cref{sec:software} describes the
\texttt{Stata} and \texttt{R} packages. Finally,
\Cref{sec:conclusions} concludes.

\section{Identification and Treatment Timing}
\label{sec:notation}

The primary objective of most analyses using the
difference-in-differences approach is to identify the average
treatment effect on the treated ($\ATT$). Following the potential
outcomes framework, the $\ATT$ is defined as:
\begin{equation}
\label{eq:defATT}
  \ATT = \E[Y_{i,r,t}(1) - Y_{i,r,t}(0)|D_r=1],
\end{equation}
where $Y_{i,r,t}(1)$ and $Y_{i,r,t}(0)$ represent potential outcomes
for person~$i$ in region~$r$ in period~$t$, with and without
treatment, respectively, and $D_r$ denotes the binary treatment
indicator, which is assigned at the region level. In the data, we only
observe realized outcomes $Y_{i,r,t}$. While \eqref{eq:defATT} defines
the target parameter, the appropriate estimator depends on the
temporal distribution of treatment across units; specifically, on
whether treatment occurs simultaneously or is staggered over time.

Consider a panel setting with regions $r \in \{1,\ldots,R\}$ observed
over time periods $t \in \{1,\ldots,T\}$. We partition regions into 
those that receive treatment ($r \in R_1$) and those that are never 
treated ($r \in R_0$). To formalize the timing of these interventions,
let $\biS = (S_1, \ldots, S_R )$ be a $1 \times R$ vector where each
entry $S_r$ denotes the first treatment period for region~$r$. For 
never-treated regions, $S_r = 0$. We assume that treatment is an
absorbing state; once a region is treated, it remains treated for all
subsequent periods $t \geq S_r$.

The complexity of the estimation strategy is determined by the number
of unique non-zero values in~$\biS$. If $\biS$ contains only a 
single unique non-zero value, then the design involves \emph{common 
adoption}. In that case, it is standard practice to use the two-way
fixed effects estimator (TWFE) discussed in \Cref{sec:twfe}. If,
however, $\biS$ contains multiple unique non-zero entries, the
design involves \emph{staggered adoption}.

In the case of staggered adoption, it is computationally and 
conceptually useful to group regions into treatment timing cohorts
indexed by~$g$. These cohorts are defined by the period of initial
treatment adoption. Let $\mathcal{G}$ be the set of unique non-zero
values in~$\biS$. For each $g \in \mathcal{G}$, we define a $1 \times
R$ binary vector~${\bm G}_g$, where the $r$-th entry $G_{g,r} = 1$ if
and only if $S_r = g$. Similarly, for the never-treated group, we
define ${\bm G}_\infty$ such that $G_{\infty,r} = 1$ if and only if
$S_r = 0$.

This cohort-based structure allows for the estimation of group-time
average treatment effects (on the treated), denoted as $\ATT(g,t)$.
These parameters represent the effect of treatment for a specific
cohort $g$ at time~$t$. For simplicity, references to the cohort 
${\bm G}_g$, or cohort~$g$, imply the set of regions with values of 1 
for that particular vector.

As discussed in \Cref{sec:cs}, the $\ATT(g,t)$ serve as the building
blocks for aggregate parameters, such as cohort-specific effects
$\ATT(g)$ or a single weighted average $\ATT$ for the entire study
period. This decomposition is particularly relevant when comparing
standard TWFE models against heterogeneous-robust estimators like those
proposed by \citet{callaway2021difference}.

\subsection{Two-Way Fixed Effects}
\label{sec:twfe}

Major innovations in DiD methods have come from taking the 
potential outcomes framework seriously and determining what assumptions
are necessary for a model to recover the relevant treatment effect. 
\citet*{Roth_2023} provides a recent survey of these developments. 
In general, DiD models are used to estimate the $\ATT$ in 
\eqref{eq:defATT} as a summary measure of treatment effects.

As only one of the potential outcomes can be realized for any
individual, care must be taken when constructing the counter-factual
potential outcome for individuals. While there are many assumptions
needed for DiD to recover the~ATT, the two most critical are the
\emph{parallel trends assumption} and the \emph{no-anticipation
assumption}. The parallel trends assumption implies that the trends of
untreated potential outcomes are parallel across treated and control
regions \citep{roth_2023PTs}, while the no-anticipation assumption
implies that treatment has no effect on treated units prior to
treatment \citep{Roth_2023}. To fix ideas, consider a simple
$2\times2$ setup where there are two regions, treatment and control,
and two time periods 1 and~2. Treatment occurs in the treated region
in period~2.

Following \citet{callaway2021difference}, under the parallel trends
and no-anticipation assumptions, the ATT in \eqref{eq:defATT} can be
expressed in terms of observable outcomes, with the unit $i$
suppressed for concision, as
\begin{equation}
\label{eq:defATTmodified}
  \ATT = \E [Y_{r,2} - Y_{r,1}|D_r=1] -
  \E [Y_{r,2} - Y_{r,1}|D_r=0].
\end{equation}
In this equation observations with $D_r=1$, or regions $r \in R_1$,
are treated, while those with $D_r=0$, or regions $r \in R_0$, are
controls. Here $Y_{r,1}$ are observations from the first time period,
and $Y_{r,2}$ are observations from the second time period.

For a long time, researchers believed that, with these two primary
assumptions satisfied, a simple DiD regression would yield the~ATT. 
With just two regions (one treatment and one control) and at least two 
time periods, the simple DiD regression is often specified as
\begin{equation}
Y_{i,r,t} = \alpha  + \beta {\rm did}_{r,t} + \gamma {\rm post}_t 
+ \lambda D_r + \epsilon_{i,r,t}.
\label{eq:2by2} 
\end{equation}
Here, ${\rm post}_t$ is an indicator variable set equal to~1 for
observations in the post intervention period, $D_r$ was previously
defined, and ${\rm did}_{r,t} = {\rm post}_t \times D_r$. The
coefficient $\beta$ on ${\rm did}_{r,t}$ will recover the ATT in this
simple setting.

In settings where there are multiple control regions and multiple 
treatment regions, it is natural to extend the simple model to
accommodate the additional regions and time periods. Specifically,
\eqref{eq:2by2} is modified by replacing ${\rm post}_t$ with a set of
year fixed effects~($\delta_t$) and $D_r$ with a set of region fixed 
effects~($\lambda_r$). This yields the TWFE model,
\begin{equation}
Y_{i,r,t} = \alpha + \beta {\rm did}_{r,t} + \delta_t + \lambda_r 
+ \epsilon_{i,r,t}.
\label{eq:TWFEreg}
\end{equation}
The ${\rm did}_{r,t}$ variable is conceptually the same as before, but
it has to be hand coded to equal~1 for treated time periods within
treated regions. This requires accounting for the treatment timing in
each region. Accordingly, ${\rm did}_{r,t}$ is set to 0 if $r\in R_0$,
and set to 1 if $r \in R_1 $ and $t \geq S_r$. Until about five years 
ago, it was believed that $\beta$ from the TWFE model would 
consistently estimate an ATT; see the discussion in the popular 
textbook \citet{MHE_2008} as an example.

In the past five or so years, however, several extremely influential
papers have shown that, among other things, the TWFE estimator is
inconsistent in some cases \citep{de2020twott,
goodman2021difference,callaway2021difference}. This failure is
particularly acute in \emph{staggered adoption} settings (where
different regions adopt policies at different times) when treatment
effects are heterogeneous across groups. This research also 
demonstrates that, with the TWFE estimator, certain treated 
\mbox{region $\times$ year} cells, which should get a positive weight 
for the average treatment effect, are instead weighted negatively. The
TWFE estimator can also include \emph{forbidden comparisons}, where 
the estimated treatment effect is a function of sub-effects calculated
based on contrasting newly treated units with previously treated ones.

\begin{example}[Staggered Adoption, $3\times3$, Single Never-Treated
Control Region]
\label{example1}
To illustrate this problem, consider the simplest $3 \times 3$ setting,
where region~A is treated in period~2, region~B is treated in period~3, 
and region~C is never treated. The adoption timing can be represented as
\begin{equation}
\label{eq:example}
  \begin{array}{c c c c}
    & {\rm A} & {\rm B} & {\rm C} \\
    1 & 0 & 0 & 0 \\
    2 & 1 & 0 & 0 \\
    3 & 1 & 1 & 0 \\
  \end{array}
\end{equation}
Let us denote, for example, a comparison between treated region~A and
not-yet (or never-treated) region~B, measured between periods~2 and~1,
as~AB21. Here period~1 is the period right before region~A is treated.
Then there are nine possible $2 \times 2$ comparisons within this $3
\times 3$ setting: AB21, AB32, AB31, AC21, AC32, AC31, BC21, BC32,
and~BC31.

Several of these $2 \times 2$ comparisons do not provide any
information, because there is no variation in treatment across regions
or time. For instance, the comparison BC21 provides no treatment information,
because no units are treated in this block. AC32 has no within-region
variation in treatment, and AB31 has no within-time variation in
treatment. The remaining six comparisons have the necessary variation
in treatment. \citet{goodman2021difference} showed that $\beta$ from
the TWFE regression in \eqref{eq:TWFEreg} is equivalent to a weighted
average of these comparisons. Specifically, AC31 and AC21 are
aggregated as ``early region vs.\ untreated region,'' and BC32 and
BC31 are regarded as ``late region vs.\ untreated region.''

The forbidden comparison here is~AB32, as this compares the late 
adopter,~B, to the early adopter,~A, when A was in fact treated in 
both periods. Both AB21 and AB31 are valid comparisons that use 
information from B \emph{before} it was treated. AC31, AC21, BC32, and
BC31 are also valid comparisons, because they use information from the
never-treated region,~C. Newer estimators have attempted to exclude
the forbidden comparisons to estimate an~ATT. Some of these use all 
pre-treatment periods, and some only use the period before
intervention. Similarly, some use the not-yet treated cells and
never-treated cells, while others exclude the not-yet treated cells.
\end{example}

\subsection{The CSDID Procedure}
\label{sec:cs}

The \citet{callaway2021difference} (CSDID) procedure, which is
implemented in the \texttt{Stata} and \texttt{Python} packages
\texttt{csdid} and the \texttt{R} package~\texttt{did}, applies a
two-step approach to estimate an~ATT. The first step estimates a
series of $2 \times 2$ treatment effects, and the second step
aggregates them. Specifically, the first step enumerates all the
two-group, two-period differences that could estimate a treatment
effect, excluding any forbidden comparisons. Each such comparison
includes a treated and a control group for one period before and one
period after treatment. The estimators focus attention on the year
immediately before treatment for all comparisons as the
pre-intervention period. The CSDID procedure then estimates a $2
\times 2$ ATT using only the observations for those four \mbox{group
$\times$ year} cells. 

In contrast to the TWFE estimator, the ``groups'' are defined by 
treatment timing, rather than by regions. These cohorts are defined
by the time when treatment was first adopted, if ever. Using the
values of $S_r$ and ${\bm G}_g$ defined in \Cref{sec:notation}, CSDID
estimates time-by-cohort treatment effects.

The $2 \times 2$ ATT based on four group-year cells is referred to as
an~$\ATT(g,t)$, where $g$ indexes the treatment timing cohort, and
$t$ indexes the post-treatment year. The control group can vary from
one $\ATT(g,t)$ to the next. It also depends on the choice of
never-treated or not-yet-treated comparison units. For a particular
$\ATT(g,t)$, we can denote the control group as~$\mathcal{G}_{{\rm
comp}(g,t)}$. It is sometimes useful to differentiate between the
never-treated control group $\mathcal{G}_{{\rm comp}(g,t),{\rm NT}}$
and the not-yet-treated control group~$\mathcal{G}_{{\rm
comp}(g,t),{\rm NY}}$. In both cases, the time periods used for the
control group are the same as for the treated group, specifically, $t$
and~$g-1$. Recall that $g-1$ is the last pre-adoption period for
cohort~$g$. 

The never-treated control group $\mathcal{G}_{{\rm comp}(g,t),{\rm
NT}}$ is comprised of the cohort~${\bm G}_\infty$, that is, the
regions with $S_r=0$. In contrast, the not-yet-treated control group 
$\mathcal{G}_{{\rm comp}(g,t), {\rm NY}}$ is comprised of the ${\bm
G}_\infty$ cohort and cohorts with $g > t$, that is, regions that were
treated after period~$t$, for which $S_r > t$. Sometimes, all groups
are eventually treated. In that case, $R_0$ is empty, and only the
not-yet-treated control group is used; see \Cref{example5} below. For
convenience, the comparison group is simply referred to
as~$\mathcal{G}_{\rm comp}$ in what follows.

Without covariates, a definition of the $\ATT(g,t)$ is
\begin{equation}
  \ATT(g,t) = \E[Y_{r,t}-Y_{r,g-1} \mid S_r = g] - 
  \E[Y_{r,t}-Y_{r,g-1} \mid r \in \mathcal{G}_{{\rm comp}}].
\label{eqn: callaway estimand}
\end{equation}
The argument $t$ is the specific post-treatment time period for which
the $\ATT(g,t)$ is calculated. For each $g \in \mathcal{G}$, there is
an $\ATT(g,t)$ for all $g \leq t \leq T$. Thus there is an $\ATT(g,t)$
for each timing cohort, for all post-treatment periods in which there
exists an untreated comparison group. The $\ATT(g,t)$ can be estimated
in various ways. If treatment is exogenous and we have a
fully-balanced panel, then a straightforward estimator, analogous to
\eqref{eqn: callaway estimand}, is
\begin{equation}
  \widehat\ATT(g,t) = \frac{1}{N_{g,t}} \sum_{r : S_r = g}
  \sum_{i= 1}^{N_{r,t}}
  (Y_{i,r,t} - Y_{i,r,g-1}) - \frac{1}{N_{\mathcal{G}_{{\rm comp},t}}}
  \sum_{r \in \mathcal{G}_{\rm comp}}\sum_{i = 1}^{N_{r,t}}
  (Y_{i,r,t} - Y_{i,r,g-1}),
\label{eqn: callaway sample analog}   
\end{equation}
where $g$ indexes the group or cohort for which the $\ATT(g,t)$ is 
estimated. This is the ${\bm G}_g$ cohort, or the cohort which is
first treated at time~$g$. Recall that this cohort is comprised of the
regions with $S_r=g$. Here, $Y_{i,r,t}$ is the post-treatment period
outcome for person $i$ in region~$r$, and $Y_{i,r,g-1}$ is the
pre-treatment period outcome. The two double summations in \eqref{eqn:
callaway sample analog} indicate that we are summing over all
observations $i$ from all regions $r$ that are part of the specified
cohort, either ${\bm G}_g$ or~$\mathcal{G}_{\rm comp}$. If
$N_{r,t}$ is the number of observations in region~$r$ in time period
$t$, then $N_{g,t}$ is the total number of observations in cohort~$g$
in time period~$t$, so that $N_{g,t} = \sum_{r : S_r = g}N_{r,t}$.
Similarly, $N_{\mathcal{G}_{{\rm comp},t}}$ is the total number of
observations in time period $t$ for the comparison group, so that
$N_{\mathcal{G}_{{\rm comp},t}} = \sum_{r \in \mathcal{G}_{\rm
comp}}N_{r,t}$.

With a fully-balanced panel, every unit is observed in every period,
and there are no missing observations. Thus $N_{g,t}$ is also the
total number of observations in cohort~$g$ in time period $g-1$. In
repeated cross-sectional datasets, or in panel datasets with
attrition, the number of observations in each cohort can differ across
time periods, and \eqref{eqn: callaway sample analog} would have to be
replaced by a more complicated equation.  

The second step involves taking a weighted average of the
$\widehat\ATT(g,t)$ to obtain an overall estimate of the~ATT,
\begin{equation}
\label{eq:widehatATT}
\widehat\ATT = \sum_{g \in \mathcal{G}} \sum_{t=g}^T w_{g,t}
\widehat\ATT(g,t).
\end{equation}
The weights $w_{g,t}$ sum to one and are chosen by the user, but the 
default is equal weighting (referred to as ``simple'' weighting), so 
that $\widehat\ATT$ is simply the average of the~$\widehat\ATT(g,t)$.

In the remainder of this subsection, we consider three examples which
illustrate how the CSDID procedure works for staggered adoption.

\begin{example}[Staggered Adoption, $3\times3$, Single Never-Treated
Control Region, CSDID]
\label{example1cs}

For the setup in  \Cref{example1}, repeated here for convenience, the
adoption timing is given by the matrix
\begin{equation*}
\tag{\ref{eq:example} repeated}
  \begin{array}{c c c c}
    & {\rm A} & {\rm B} & {\rm C} \\
    1 & 0 & 0 & 0 \\
    2 & 1 & 0 & 0 \\
    3 & 1 & 1 & 0 \\
  \end{array}
\end{equation*}
Using the above notation, region~A belongs to cohort $g=2$, region~B
belongs to cohort $g=3$, and region~C belongs to cohort $g=\infty$.
The set of treated cohorts is thus $\mathcal{G}=\{2,3\}$.

For this example, there are three $\ATT(g,t)$ terms to estimate, 
$\ATT(2,2)$, $\ATT(2,3)$, and~$\ATT(3,3)$. In principle, we could
compare the treated units either with never-treated units or with
not-yet-treated units. By default, the \texttt{csdid} package uses
only never-treated units when that is feasible. In this case,
$\mathcal{G}_{{\rm comp}(2,2),{\rm NT}} = \mathcal{G}_{{\rm
comp}(2,3),{\rm NT}} = \mathcal{G}_{{\rm comp}(3,3),{\rm NT}} = \{C\}$.
Thus the relevant comparisons are AC21, AC31, and~BC32. These are used 
to estimate $\ATT(2,2)$, $\ATT(2,3)$, and $\ATT(3,3)$, respectively, 
and the overall estimand of interest is
\begin{equation}
  \ATT = \frac{1}{3}\big(\ATT(2,2) + \ATT(2,3) + \ATT(3,3)\big).
\end{equation}
The individual $\ATT (g,t)$ components are
\begin{align}
\label{indivATT}
  \ATT(2,2) &= \big(\E [Y_{{\rm A},2} - Y_{{\rm A},1}]\big) -
  \big(\E[Y_{{\rm C},2} - Y_{{\rm C},1}]\big),\notag \\
  \ATT(2,3) &= \big(\E[Y_{{\rm A},3} - Y_{{\rm A},1}]\big) -
  \big(\E[Y_{{\rm C},3} - Y_{{\rm C},1}]\big), \\
  \ATT(3,3) &= \big(\E[Y_{{\rm B},3} - Y_{{\rm B},2}]) -
  \big(\E[Y_{{\rm C},3} -Y_{{\rm C},2}]\big).\notag
\end{align}
Here, $\E [Y_{{\rm A},2}]$ is the expected value of the outcome
variable for all individuals~$i$ in region~A in period~2, and the
other expectations are defined similarly. When using the default
``simple'' aggregation weights, the \texttt{csdid} package estimates
the ATT using
\begin{equation}
  \widehat\ATT = \frac{1}{3}\big(\widehat\ATT(2,2) + \widehat\ATT(2,3) 
  + \widehat\ATT(3,3)\big).
\label{eq:cssampleagg}
\end{equation}
Each of the $\widehat\ATT (g,t)$ terms is first estimated 
using~\eqref{eqn: callaway sample analog}. Without covariates, a 
numerically identical estimate of the ATT can be obtained by using 
three individual $2 \times 2$ DiD models like~\eqref{eq:2by2}.

It is possible to use not-yet-treated units instead of never-treated
units as controls, even when the latter is feasible, by specifying the
\texttt{notyet} option. This may provide more efficient estimates of
the ATT, but it requires a no-anticipation assumption for those units.
For this example, $\mathcal{G}_{{\rm comp}(2,2),{\rm NY}} = 
\{{\rm B},{\rm C}\}$ and $\mathcal{G}_{{\rm comp}(2,3),{\rm NY}} = 
\mathcal{G}_{{\rm comp}(3,3),{\rm NY}}= \{\rm{C}\}$. Thus the only 
difference between the NY and NT control groups is that the former 
includes region~B for estimating~$\ATT(2,2)$. The first line in 
\eqref{indivATT} would therefore be replaced by
\begin{equation}
  \ATT(2,2) = \big(\E[Y_{{\rm A},2} - Y_{{\rm A},1}]\big) -
  \big(\omega(\E[Y_{{\rm B},2} - Y_{{\rm B},1}]) + (1-\omega)
  (\E[Y_{{\rm C},2} - Y_{{\rm C},1}])\big),
\label{ATT22}
\end{equation}
where $\omega$ is a weight. It could be 1/2, or it could equal the
average number of individuals in region~B in the two periods divided
by the average number in both regions~B and~C. The other two lines
would be unchanged.
\end{example}

\begin{example}[Staggered Adoption, $6\times3$, Two Never-Treated Control
Regions]
\label{example_extended}

Now suppose we double the number of regions. Specifically, 
there are $R=6$ regions, $r \in \{A,B,C,D,E,F\}$, and $T=3$ 
periods, $t\in\{1,2,3\}$. The treatment adoption schedule is
\begin{equation}
  \label{eq:example_extended}
    \begin{array}{c c c c c c c}
      & {\rm A} & {\rm B} & {\rm C} & {\rm D} & {\rm E} & {\rm F} \\
      1 & 0 & 0 & 0 & 0 & 0 & 0 \\
      2 & 1 & 0 & 0 & 1 & 0 & 0 \\
      3 & 1 & 1 & 0 & 1 & 1 & 0 \\
  \end{array}
\end{equation}
In this case, $R_1=\{{\rm A},{\rm B},{\rm D},{\rm E}\}$ and
$R_0=\{{\rm C},{\rm F}\}$. For this example, the timing vector $\biS$
(first treatment period by region) is \[ \biS=(S_{\rm A},S_{\rm
B},S_{\rm C},S_{\rm D},S_{\rm E},S_{\rm F})=(2,3,0,2,3,0). \] For
reference, these $S_r$ values can be represented in a matrix
aligned with periods~$t$ as follows
\begin{equation}
  \label{tab:example_timing}
    \begin{array}{c c c c c c c}
      & {\rm A} & {\rm B} & {\rm C} & {\rm D} & {\rm E} & {\rm F} \\
      1 & 2 & 3 & 0 & 2 & 3 & 0 \\
      2 & 2 & 3 & 0 & 2 & 3 & 0 \\
      3 & 2 & 3 & 0 & 2 & 3 & 0 \\
    \end{array}
\end{equation}
The values repeat by row because $S_r$ is time-invariant. For those
familiar with the software packages (\texttt{csdid}, etc.), note that
they require an option called~\texttt{gvar}. This is an
observation-level variable that is constant within region. Specifically,
\texttt{gvar} that takes the value $S_r$ for every individual~$i$ in 
region~$r$.

There are three cohorts: ${\bm G}_2$, ${\bm G}_3$, and~${\bm G}_\infty$.
Their indicators are
\begin{equation}
\label{eq:example_cohorts}
  \begin{array}{l c c c c c c}
    & {\rm A} & {\rm B} & {\rm C} & {\rm D} & {\rm E} & {\rm F} \\
    {\bm G}_2 & 1 & 0 & 0 & 1 & 0 & 0 \\
    {\bm G}_3 & 0 & 1 & 0 & 0 & 1 & 0 \\
    {\bm G}_\infty & 0 & 0 & 1 & 0 & 0 & 1 \\
  \end{array}
\end{equation}
Accordingly, $\mathcal{G}=\{2,3\}$.

As an illustration, consider $\ATT(3,3)$. The treated cohort is ${\bm
G}_3$ (regions $B$ and~$E$). Using never-treated units, the 
comparison group is $\mathcal{G}_{{\rm comp}(3,3),{\rm NT}} = \{{\rm
C},{\rm F}\}$. The two periods are $g-1=2$ (pre) and $t=3$ (post).
Thus
\begin{align*}
   \ATT(3,3) &= \big(\omega_1(\E[Y_{{\rm B},3} - Y_{{\rm B},2}])
   + (1-\omega_1)(\E[Y_{{\rm E},3} - Y_{{\rm E},2}])\big)\\
   &\phantom{=}\;- \big(\omega_2(\E[Y_{{\rm C},3} - Y_{{\rm C},2}]) + 
   (1-\omega_2)(\E[Y_{{\rm F},3} - Y_{{\rm F},2}])\big),
\end{align*}
where $\omega_1$ and $\omega_2$ are weights, like the one in
\eqref{ATT22}.

Without covariates, $\ATT(3,3)$ can be estimated using a conventional
DiD model with two cohorts and two time periods. Define the sample to
be observations in cohorts ${\bm G}_3$ or ${\bm G}_\infty$ \emph{and} 
in periods 2 or~3. Then estimate the following $2\times2$ DID model 
using this sample:
\begin{equation}
  Y_{i,g,t} = \alpha + \delta\tk\IF(t=3) + \lambda\tk\IF(g=3) + \beta 
  (\IF(t=3)\times\IF(g=3) ) + \epsilon_{i,r,t},
\label{eq:2by2attgt}
\end{equation}
where $\IF(\cdot)$ is the indicator function. In this regression,
$\hat \beta$ estimates~$\ATT(3,3)$.
\end{example}

\begin{example}[Staggered Adoption, $3\times4$, All Regions Treated]
\label{example5}

Now consider adding a fourth period to \Cref{example1}, in which region C 
is treated for the first time. With this addition, the adoption timing
is now given by the matrix
\begin{equation}
\label{eq:examplent}
  \begin{array}{c c c c}
   & {\rm A} & {\rm B} & {\rm C}\\
     1 & 0 & 0 & 0 \\
     2 & 1 & 0 & 0 \\
     3 & 1 & 1 & 0 \\
     4 & 1 & 1 & 1 \\
  \end{array}
\end{equation}
In this case, $R_1 = \{{\rm A},{\rm B},{\rm C}\}$, and $R_0 =
\emptyset$. As a result, the never-treated option is infeasible, and
the not-yet-treated set of comparison groups has to be used. The
timing vector is
\begin{equation*}
\biS=(S_{\rm A},S_{\rm B},S_{\rm C})=(2,3,4).
\end{equation*}
Here, despite the fourth period, it is still only possible to estimate
$\ATT(2,2), \ATT(2,3)$, and~$\ATT(3,3)$. We cannot estimate any
$\ATT(g,4)$ because there are no untreated units in the fourth period.
The comparison groups for the $\ATT (g,t)$ that we can estimate are
$\mathcal{G}_{{\rm comp}(2,2),{\rm NY}} = \{{\rm B},{\rm C}\}$, 
$\mathcal{G}_{{\rm comp}(2,3),{\rm NY}} = 
\mathcal{G}_{{\rm comp}(3,3),{\rm NY}} = \{{\rm C}\}$.
\end{example}

\section{Cluster-Robust Inference for CSDID}
\label{sec:cluster}

Clustering at the region level causes difficulties for inference. The
estimated ATT is normally a weighted sum of several
$\widehat\ATT(g,t)$ terms; see, for example,~\eqref{eq:cssampleagg}.
If we want to cluster at the region level, we have to take into
account correlations of the scores across all observations that are
used to estimate the~ATT. These correlations arise because
observations from a single region may be used to estimate multiple
$\ATT(g,t)$ terms. For instance, in \Cref{example1}, A,1 and C,1 are
used in the estimation of both $\ATT(2,2)$ and~$\ATT(2,3)$, and C,3 is
used in the estimation of both $\ATT(2,3)$ and~$\ATT(3,3)$. This
implies that data from region~C are used to estimate all three
$\ATT(g,t)$ terms, and that data from region~A are used to estimate
both $\ATT(2,2)$ and~$\ATT(2,3)$. A valid routine for cluster-robust
inference has to account for the resulting correlations.

Difficulties with inference when the observations are clustered is not
unique to the procedures of \citet{callaway2021difference}, as two
recent papers discuss. \citet{mizushima2025inference} investigates the
size properties of several modern DiD estimators. It finds that the
default, or asymptotic, procedures are often over-sized when there are
few treated clusters. However, the wild cluster bootstrap and a
randomization inference procedure can both greatly improve test size
for imputation-type tests. \citet{weiss2024much} considers the
inference properties of many DiD estimators and finds that all of them
can lack power, notably CSDID, TWFE, imputation
\citep*{borusyak2024revisiting}, and DCDH \citep{de2020twott}. In fact, 
power can even decline when additional time periods are included in the
analysis.

Neither of these papers considers the jackknife for improving
inference, as we do in \Cref{sec:jack}, although \citet{weiss2024much}
does consider the so-called~$\mbox{CV}_{\tn2}$ variance estimator; see
\citet[Section~2.2]{MNW-guide}.

\subsection{Recentered Influence Function (RIF) Standard Errors}
\label{sec:se}

The default method for obtaining standard errors in
\citet{callaway2021difference} uses the contributions to the influence
functions by each of the observations. The expressions for the
influence functions are quite long, so we omit them; see Theorem~2 of
the paper. Here we are concerned not with how the RIF variance
estimator is derived but with how well it performs in practice.

A simplified version of the second result in
\citet[Theorem~2]{callaway2021difference} can be written as
\begin{equation}
\sqrt{n} \left(\widehat{\textbf{ATT}}_{g\leq t}
- \textbf{ATT}_{g\leq t}\right) \dto \N(\bzero,\bSigma).
\label{ATTvar}
\end{equation}
Here, following the notation in \cite{santanna2023did},
$\textbf{ATT}_{g \leq t}$ and $\widehat{\textbf{ATT}}_{g \leq t}$ denote
the vectors of $\ATT(g, t)$ and $\widehat\ATT(g, t)$, respectively,
for all $g = 2, \ldots, t$ and $t = 2, \ldots, T$. The variance matrix 
$\bSigma$ is defined by
\begin{equation*}
\bSigma = \E[\bPsi_{g\leq t}(\biW_i) \bPsi_{g\leq t}(\biW_i)^\top],
\end{equation*}
where $\biW_i$ is the data vector for observation~$i$, and 
$\bPsi_{g\leq t}(\biW_i)$ are the influence functions. The result 
\eqref{ATTvar} is used to compute asymptotic RIF standard errors based 
on $\hat\bSigma$, the sample analog of $\bSigma$. Note that 
constructing $\hat\bSigma$ can be a bit tricky.

These RIF standard errors are then used with the standard normal
distribution to calculate $P$~values and confidence intervals. This
makes sense when there is no clustering, which is assumed, as the
presence of the factor $\sqrt{n}$ in~\eqref{ATTvar} makes evident.
When the observations are clustered, however, it seems like an odd
thing to do. Almost all empirical work that uses cluster-robust
$t$-statistics for linear regression models assumes that these are
distributed as $t(H-1)$, where $H$ is the number of clusters. As
\citet*{MNW-logit} illustrates, using the $\N(0,1)$ distribution
instead of the \mbox{$t(H-1)$} distribution increases rejection
frequencies substantially when $H$ is small. Thus, it seems highly
unlikely that the default RIF method for inference will yield reliable
results when there is clustering and $H$ is not large. Indeed,
\citet[Remark~13]{callaway2021difference} points out that their
approach requires a large number of clusters.

\subsection{The Multiplier Bootstrap}
\label{sec:boot}

As an alternative way to obtain standard errors, which may or may not
be cluster-robust, \citet{callaway2021difference} suggests using a
form of multiplier bootstrap. For each of $B$ bootstrap samples,
indexed by~$b$, a vector of bootstrap estimates is constructed as
\begin{equation}
\widehat{\textbf{ATT}}^*_{b,(g\leq t)} = \widehat{\textbf{ATT}}_{g\leq t} 
+ \frac{1}{n} \sum_{i=1}^n V_{b,i}^* \hat\bPsi_{g\leq t}(\biW_i).
\label{ATTboot}
\end{equation}
Here $\hat\bPsi_{g\leq t}(\biW_i)$ is the sample analog of the
influence function $\bPsi_{g \leq t}(\biW_i)$. The $V_{b,i}^*$ are
random draws from an auxiliary distribution with mean~0 and
variance~1, typically the Mammen distribution. When the data are
clustered, these draws are constant within each cluster, as for the
wild cluster bootstrap; see \citet{santanna2023did} for additional
details. The bootstrap estimates are centered around the original
estimate, with perturbations of the influence functions at the
observation or cluster level added to them. The $B$ vectors of
bootstrap $\widehat\ATT^*_b(g,t)$ values given by \eqref{ATTboot} are 
then used to form confidence intervals for the individual $\ATT(g,t)$ 
or for the~ATT.

This multiplier bootstrap is conceptually simple and should be
computationally efficient if it is implemented appropriately. Using it
avoids the need to construct $\hat\bSigma$ analytically. It differs
from the widely used wild cluster bootstrap \citep*{CGM_2008,DMN_2019}
in one important way. With the wild cluster bootstrap, each bootstrap
replication involves re-estimating the coefficients and test
statistics of interest. With the multiplier bootstrap, however, these
are not re-estimated. The bootstrap draws simply add noise to the
original estimate, and the $\widehat\ATT^*_b(g,t)$ that result are
then used to estimate how much the $\widehat\ATT(g,t)$ vary. Since the
model is never re-estimated, this is just a convenient way to 
estimate~$\bSigma$. Thus it is not surprising that the multiplier 
bootstrap and asymptotic approaches to inference yield very similar 
answers; see \Cref{sec:MCR}.

\section{The Cluster Jackknife}
\label{sec:jack}

As an alternative to the methods suggested in
\citet{callaway2021difference}, we propose to use the cluster
jackknife to calculate standard errors for~$\widehat\ATT$. The cluster
jackknife has been shown to perform well in many cases, especially in
settings with few and unbalanced clusters
\citep*{MNW-bootknife,Hansen_2025}. Another attractive feature of the
jackknife is that no analytical expressions for the influence
functions are needed. Instead, the cluster jackknife simply omits each
cluster $h$ in turn, estimates each of the $\ATT(g,t)$, and then uses
those estimates to re-estimate the~ATT. The standard error is then
simply the square root of
\begin{equation}
\mbox{CV}_{\tn3}(\widehat\ATT) = \frac{H-1}{H} \sum_{h=1}^H
\big(\widehat\ATT^{(h)} - \widehat\ATT \big)^2 , 
\end{equation}
where $\widehat\ATT$ is the original CSDID estimate of the~ATT, and
$\widehat\ATT^{(h)}$ is the estimate when cluster~$h$ is omitted.
Here, following much of the literature, we use CV$_{\tn3}$ to denote
the cluster-jackknife variance estimator, because it generalizes the
jackknife HC$_3$ variance estimator of \citet{MW_1985}.

Although the cluster jackknife has proven to work well in many
settings, a few caveats are in order. The first is that the jackknife
is not amenable to all staggered adoption settings.
\citet{weiss2024much} discusses this issue and does not investigate
the jackknife for that reason. For instance, consider \Cref{example1}
with clustering at the region level. In this case, some of the
$\ATT(g,t)$ terms cannot be calculated when certain regions are
dropped. The $\ATT(2,2)$ and $\ATT(2,3)$ terms cannot be calculated
when region~A is dropped, the $\ATT(3,3)$ term cannot be calculated
when region~B is dropped, and no ATT terms can be calculated when
region~C is dropped. In contrast, all of the $\ATT(g,t)$ terms can be
estimated for each jackknife sample for a setup like the one in
\Cref{example_extended}. For that reason, we designed the Monte Carlo
simulations described in \Cref{sec:MCD} so that there are always
multiple treated clusters and untreated clusters. However, in simulations
with one early adopter and one late adopter, there are not multiple regions
in each cohort. 

The second caveat is that some staggered adoption settings will alter 
the weights given to different $\ATT(g,t)$ across jackknife 
replications. Consider a modification of \Cref{example1}, where there 
are now two control groups, C and~D. In this case, it would be
possible to estimate an ATT for each jackknife replication. Dropping
either region~C or~D would still allow all three $\ATT(g,t)$ terms to
be calculated. However, dropping region~A would mean that the ATT
would be based solely on~$\ATT(3,3)$. Likewise, dropping region~B
would mean that the ATT would just be a weighted average of
$\ATT(2,2)$ and~$\ATT(2,3)$. If the ATT is sensitive to the regions
used to compute it, this variation in weights may highlight that fact
and thus provide valuable information. However, the
$\widehat\ATT^{(h)}$ may now vary across jackknife samples in ways
that would not happen for other applications of the jackknife. Thus,
unlike with other inference procedures, care must be taken to assess
whether the jackknife is a sensible tool to use for a given treatment
design.

The third caveat is that the jackknife procedure can be slow. Many of
the modern DiD routines, including CSDID, can be slow to calculate
when samples are large and there are many $\ATT(g,t)$ terms to
estimate. The jackknife we propose needs to call whatever command
implements CSDID $H+1$ times, once for the full sample estimate
$\widehat\ATT$, and $H$ additional times for each of the
$\widehat\ATT^{(h)}$ estimates. Fortunately, the results of our
simulations in \Cref{sec:MCR} suggest that this increase in
computational time can result in large reductions in over-rejection by
hypothesis tests or under-coverage by confidence intervals.

\section{Monte Carlo Design}
\label{sec:MCD}

Our Monte Carlo simulations use Current Population Survey (CPS) data.
The objective is to compare the rejection frequencies for several
different tests of $\ATT=0$. We conduct a set of ``placebo laws''
experiments inspired by \citet*{BDM_2004}, in which we pretend that
treatment occurred in various states at various times. Independently,
\citet{mizushima2025inference} ran very similar experiments, comparing
several DiD estimators. Their simulations examine state-by-year 
unemployment rates, while ours investigate earnings at the individual
level. Our simulations were conducted in \texttt{Stata}, using both
the \texttt{csdid} package and our new \texttt{csdidjack} package. In
our experiments, we compare the rejection frequencies using the
default \texttt{cluster} option and the bootstrap option
\texttt{wboot} within \texttt{csdid} with those using jackknife
standard errors from our new post-estimation package
\texttt{csdidjack} discussed in \Cref{sec:software}.

We employ the same dataset used for the placebo-laws simulations in
\cite{MW-JAE}. Our dataset is extracted from the Merged Outgoing 
Rotation Group (MORG) of the CPS. It includes individual-level 
information on weekly earnings, state, and year, covering the period 
from 1979 to~1999. The dependent variable is the log of women's 
earnings. We exclude observations with earnings less than \$20, which 
may be erroneous. The total sample, from all years and states, has 
547,818 observations. Across replications, we randomly pick 8 
consecutive years of data and $R$ states from the full sample. To
mimic staggered adoption, treatment is assigned in two cohorts, an
early cohort of $J$ states treated in year~4, and a late cohort of $L$
states treated in year~6. This results in $J+L$ total treated states,
and we set $J=L$. 

Note that treatments apply to all individuals in a state. Treatment is
also ``absorptive,'' so that once a state is treated in a particular
replication, it is always treated. For the TWFE estimator, there can
be severe inferential problems when there are few clusters or few
treated clusters. To see whether similar problems arise for the CSDID
procedure, we consider different values for both the number of states
$R$ and the numbers of early and late adopters. \Cref{tab:one} shows
the various values of $R$, $J$, and $L$ that we consider.

\begin{table}[t]
\centering
\caption{Values of $R$ and $J=L$ studied in our experiments}
\label{tab:one}
\vskip -6pt
\begin{tabular*}{0.8\textwidth}{@{\extracolsep{\fill}} l cccc }
\toprule
& $R=8$ & $R=16$ & $R=24$ & $R=32$ \\
\midrule
$J=L=1$  & \checkmark & \checkmark & \checkmark & \checkmark \\
$J=L=2$  & \checkmark & \checkmark & \checkmark & \checkmark \\
$J=L=3$  & \checkmark & \checkmark & \checkmark & \checkmark \\
$J=L=4$  &            & \checkmark & \checkmark & \checkmark \\
$J=L=6$  &            &            & \checkmark & \checkmark \\
$J=L=8$  &            &            & \checkmark & \checkmark \\
$J=L=10$ &            &            &            & \checkmark \\
$J=L=12$ &            &            &            & \checkmark \\
\bottomrule
\end{tabular*}
\end{table}

In each simulation, after obtaining a subsample, we estimate the ATT
using the \texttt{csdid} package, both with and without covariates. 
In all simulations, we use simple aggregation (i.e. unweighted
averaging) to estimate the~ATT, and we test the hypothesis that
$\ATT=0$. The covariates are age, age squared, and a set of four
education dummies. These covariates, and the outcome variable, have
all been demeaned at the state level using the full set of data. This
demeaning should strengthen the null of no treatment effects.

We calculate three rejection frequencies at the 5\% level, always
clustering at the state level. The first is for $t$-statistics based
on the ``cluster'' option of \texttt{csdid}, which uses the recentered
influence functions discussed in \Cref{sec:se}. The second is for the
multiplier bootstrap discussed in \Cref{sec:boot}. Since the bootstrap
procedure in \texttt{csdid} does not provide a $P$~value, we calculate
the rejection rate based on whether the 95\% confidence interval
covers the true value of~0. The final rejection frequency is for
$t$-statistics based on cluster-jackknife standard errors discussed in
\Cref{sec:jack} and calculated using our \texttt{csdidjack} package.
Since all three inference methods use states as clusters, $H=R$ in all
our experiments.

Our experiments have 2,400 replications. This number is quite a bit
smaller than we would like, but the simulations are very slow,
particularly for the jackknife. We hope to speed them up in the future
using a methodology similar to the one in the \texttt{fastdid} package
in~\texttt{R}. This package is up to 100x faster than the 
\texttt{did} package in~\texttt{R}, but it requires panel data rather
than cross-sectional data. It can be found at
\url{https://cran.r-project.org/web/packages/fastdid/index.html}.

\section{Monte Carlo Results}
\label{sec:MCR}

\Cref{tab:rejection-cov} shows rejection frequencies for the Monte
Carlo simulations with covariates. There are three panels, one for
each of the inference methods that we study. The results for the 
two procedures proposed in \citet{callaway2021difference} in Panels~A
and~B generally resemble each other greatly, while those for our 
cluster-jackknife procedure in Panel~C are always much closer to the 
nominal level of~0.05.

\begin{table}[tp]
\caption{Rejection frequencies for models with covariates}
\label{tab:rejection-cov}
\vskip -6pt
\begin{tabular*}{\textwidth}{@{\extracolsep{\fill}} l cccc }
\toprule
& $H=R=8$ & $H=R=16$ & $H=R=24$ & $H=R=32$ \\
\midrule
\multicolumn{5}{l}{Panel A: Asymptotic (RIF) using \texttt{csdid}} \\
\midrule
$J=L=1$  & 0.3046 & 0.3333 & 0.3558 & 0.3800 \\
$J=L=2$  & 0.1779 & 0.1867 & 0.1750 & 0.1979 \\
$J=L=3$  & 0.2146 & 0.1245 & 0.1371 & 0.1388 \\
$J=L=4$  &        & 0.1132 & 0.1204 & 0.1187 \\
$J=L=6$  &        &        & 0.0829 & 0.0954 \\
$J=L=8$  &        &        & 0.0950 & 0.0850 \\
$J=L=10$ &        &        &        & 0.0779 \\
$J=L=12$ &        &        &        & 0.0925 \\
\midrule
\multicolumn{5}{l}{Panel B: Multiplier bootstrap using \texttt{csdid}} \\
\midrule
$J=L=1$  & 0.3079 & 0.3458 & 0.3725 & 0.3950 \\
$J=L=2$  & 0.1808 & 0.1887 & 0.1762 & 0.2004 \\
$J=L=3$  & 0.2125 & 0.1233 & 0.1408 & 0.1388 \\
$J=L=4$  &        & 0.1161 & 0.1200 & 0.1204 \\
$J=L=6$  &        &        & 0.0846 & 0.0967 \\
$J=L=8$  &        &        & 0.0929 & 0.0842 \\
$J=L=10$ &        &        &        & 0.0787 \\
$J=L=12$ &        &        &        & 0.0938 \\
\midrule
\multicolumn{5}{l}{Panel C: Cluster jackknife using \texttt{csdidjack}} \\
\midrule
$J=L=1$  & 0.0938 & 0.1279 & 0.1450 & 0.1642 \\
$J=L=2$  & 0.0379 & 0.0763 & 0.0796 & 0.0929 \\
$J=L=3$  & 0.0533 & 0.0488 & 0.0600 & 0.0688 \\
$J=L=4$  &        & 0.0505 & 0.0621 & 0.0571 \\
$J=L=6$  &        &        & 0.0450 & 0.0571 \\
$J=L=8$  &        &        & 0.0425 & 0.0483 \\
$J=L=10$ &        &        &        & 0.0450 \\
$J=L=12$ &        &        &        & 0.0500 \\
\bottomrule
\end{tabular*}
\vskip 4pt 
{\footnotesize \textbf{Notes:} Nominal level is 5\%.}
\end{table}

For the asymptotic and multiplier bootstrap methods, there are clearly
severe problems with inference when there are few clusters. The ideal
case for each value of $H$ is when half the clusters are treated; see
\citet{MW-EJ} and many other papers. This case occurs when
$H=2(J+L)=4J$. Except with $H=32$, the methods in Panels~A and~B work
best for that case. For $H=32$, they work slightly better when $J=10$
than when $J=8$, but that may be due to the small number of
replications. It is clear that both methods improve as $H$ increases,
but they never perform very well. In contrast, the jackknife, in
Panel~C, always rejects around 5\% of the time for several of the
better cases.

For every method, there are serious inferential problems when there
are few treated clusters. The top row of each panel, where $J=L=1$,
always contains the largest rejection frequencies. The asymptotic and
bootstrap procedures reject between 30\% and 40\% of the time. In
contrast, the jackknife procedure never rejects much more than 16\% of
the time. Increasing the number of early and late adopters reduces
rejection rates substantially, at least up to the point where $H=4J$.

The jackknife is always reliable when $J=L$ is large enough.
Interestingly, the number of treated clusters that are needed for the
jackknife to be reliable depends on the total number of clusters.
Similar results were found for TWFE with CV$_1$ inference in
\citet{MW-JAE} and with jackknife inference in \citet*{MNW-bootknife}.

The similarity of the asymptotic and bootstrap rejection frequencies
is immediately evident from the table. At most, they differ by about
2\% (0.3333 vs 0.3458), but by much smaller amounts when the magnitude
of the over-rejection is smaller. It is clear that the multiplier
bootstrap does not improve inference much at all, as the discussion in
\Cref{sec:boot} suggested. For that reason, we will not focus on the
multiplier bootstrap in the empirical examples.

\Cref{tab:rejection-nocov} shows rejection frequencies for the
regressions without covariates. These results are broadly similar to
the results with covariates. Again, the asymptotic and bootstrap
tests are always very similar and always over-reject, in some cases
very severely. The jackknife tests generally provide much more
reliable inferences, but they can still over-reject substantially when
there are few treated clusters.

\begin{table}[tp]
\caption{Rejection frequencies for models without covariates}
\label{tab:rejection-nocov}
\vskip -6pt
\begin{tabular*}{\textwidth}{@{\extracolsep{\fill}} l cccc }
\toprule
& $H=R=8$ & $H=R=16$ & $H=R=24$ & $H=R=32$ \\
\midrule
\multicolumn{5}{l}{Panel A: Asymptotic (RIF) using \texttt{csdid}} \\
\midrule
$J=L=1$ & 0.3150 & 0.3383 & 0.3683 & 0.3725 \\
$J=L=2$ & 0.1717 & 0.1937 & 0.1813 & 0.2042 \\
$J=L=3$ & 0.2233 & 0.1279 & 0.1392 & 0.1504 \\
$J=L=4$ &        & 0.1144 & 0.1225 & 0.1208 \\
$J=L=6$ &        &        & 0.0892 & 0.0975 \\
$J=L=8$ &        &        & 0.0929 & 0.0821 \\
$J=L=10$ &       &        &        & 0.0763 \\
$J=L=12$ &       &        &        & 0.0921 \\
\midrule
\multicolumn{5}{l}{Panel B: Multiplier bootstrap using \texttt{csdid}} \\
\midrule
$J=L=1$ & 0.3171 & 0.3492 & 0.3808 & 0.3879 \\
$J=L=2$ & 0.1750 & 0.1954 & 0.1821 & 0.2079 \\
$J=L=3$ & 0.2208 & 0.1308 & 0.1396 & 0.1483 \\
$J=L=4$ &        & 0.1165 & 0.1258 & 0.1192 \\
$J=L=6$ &        &        & 0.0887 & 0.0983 \\
$J=L=8$ &        &        & 0.0938 & 0.0829 \\
$J=L=10$ &       &        &        & 0.0804 \\
$J=L=12$ &       &        &        & 0.0950 \\
\midrule
\multicolumn{5}{l}{Panel C: Cluster jackknife using \texttt{csdidjack}} \\
\midrule
$J=L=1$ & 0.0942 & 0.1279 & 0.1475 & 0.1550 \\
$J=L=2$ & 0.0358 & 0.0733 & 0.0775 & 0.1013 \\
$J=L=3$ & 0.0571 & 0.0505 & 0.0646 & 0.0679 \\
$J=L=4$ &        & 0.0538 & 0.0662 & 0.0658 \\
$J=L=6$ &        &        & 0.0487 & 0.0546 \\
$J=L=8$ &        &        & 0.0467 & 0.0475 \\
$J=L=10$ &       &        &        & 0.0442 \\
$J=L=12$ &       &        &        & 0.0575 \\
\bottomrule
\end{tabular*}
\vskip 4pt 
{\footnotesize \textbf{Notes:} Nominal level is 5\%.}
\end{table}

Overall, there are four major takeaways from these experiments:

\begin{enumerate}

\item The CSDID estimator with asymptotic or multiplier bootstrap 
standard errors is prone to over-reject, often very severely, when 
there are few clusters.

\item The CSDID estimator with asymptotic or multiplier bootstrap 
standard errors is especially prone to over-reject when there are few 
treated clusters.

\item The asymptotic and multiplier bootstrap standard errors proposed
in \citet{callaway2021difference} are generally extremely similar for
inference about the~ATT, as the discussion in \Cref{sec:boot}
suggests.

\item Inference based on cluster-jackknife standard errors is always
much better than inference based on asymptotic or multiplier bootstrap
standard errors. However, there can still be noticeable over-rejection
when the number of treated clusters is small relative to the total
number of clusters.

\end{enumerate}

\section{Empirical Examples}
\label{sec:examples}

We now present two brief empirical examples to demonstrate that there
can be meaningful differences between inferences based on asymptotic
and cluster-jackknife standard errors. Both examples use American data
clustered by state. The first example uses individual-level data, and the
second uses county-level data. Both population and the number of
counties vary considerably across states.

\subsection{Prenatal Substance Use Policies and Mental Health}
\label{subsec:substance}

We first consider an empirical example from \citet*{flex_2026}, which
uses the variation explored in \citet*{meinhofer2022prenatal}. The
latter paper examines whether the enactment of punitive prenatal
substance-use policies affects outcomes of children, while the former
considers whether these policies had an impact on the mental health of
mothers of child-bearing age. We thank the authors of
\citet{flex_2026} for making their replication files easily available.

We use data from the Behavioral Risk Factor Surveillance System
(BRFSS) for 2005--2018, a large annual survey conducted by the Centers
for Disease Control and Prevention. We restrict attention to women
aged 18--44 with at least one child, resulting in a sample of 440,446
observations from 34 U.S.\ states. Our outcome of interest is the
number of days in the past month that a respondent reports having good
mental health. This variable is highly left-skewed, with a sample mean
of approximately 25.5 days.

Thirteen states adopted punitive prenatal substance-use policies
during the sample period. The timing of adoption varies considerably,
with Idaho adopting in 2007, South Carolina in 2008, Arizona in 2009,
and a number of states enacting policies in 2012--2018. The
never-treated group consists of 21 states that never adopted such
policies during the sample period.

The analysis includes a limited set of covariates: age, race (white,
black, hispanic), education (no degree, high school, some college,
college degree), and the log of income. The mean age of respondents is
34 years.

We estimate the ATT for the CSDID estimator with simple aggregation
and clustered at the state level, using the \texttt{csdid} package. As
shown in \Cref{tab:att-results}, this standard approach yields a
statistically significant estimate at the 5\% level. However, the
cluster jackknife (using \texttt{csdidjack}) leads to a larger
standard error and a $P$~value greater than~0.05. The sample is
relatively unbalanced across the 34 states, with the largest states
having nearly 5 times as many observations as the smallest ones. This
is a setting in which we might expect the jackknife to be more
reliable \citep*[Section~6]{MNW-bootknife}.

\begin{table}[tp]
\centering
\caption{Estimates of ATT for good mental health days}
\label{tab:att-results}
\vskip -6pt
\begin{tabular*}{\textwidth}{@{\extracolsep{\fill}}
  l d{2.3} d{2.3} d{2.3} d{2.3} d{2.3} }
\toprule
Method & \multicolumn{1}{c}{ATT} & \multicolumn{1}{c}{Std.\ error}
  & \multicolumn{1}{c}{$P$ value} & \multicolumn{1}{c}{95\% CI lower}
  & \multicolumn{1}{c}{95\% CI upper} \\
\midrule
CSDID asymptotic     &-0.212 &0.102 &0.038 &-0.412 &-0.011 \\
Cluster jackknife (CV$_{\tn3}$) &-0.212 &0.113 &0.069 &-0.441 &0.017 \\
\bottomrule
\end{tabular*}
\end{table}

\subsection{Minimum Wage and Teen Employment}
\label{wage}

We also consider an empirical example adapted from 
\citet{callaway2021difference}. The analysis examines the impact of
minimum-wage increases on teen employment using county-level data from
the Quarterly Workforce Indicators for 2001--2007. During this time, 
the federal minimum wage remained flat, allowing us to define treatment 
based on when a state first increased its minimum wage. We focus on 29
states that changed their minimum wage during the sample period. The 
outcome is the log of teen employment (\texttt{lemp}), and the 
covariates include demographic and economic controls: region
indicators, percent white, percent with a high school diploma, poverty
rate, and both county population and median income as quadratics.

Estimation is performed using the \texttt{csdid} command with simple 
aggregation, clustered at the state level. Results are in Panel~A of 
\Cref{tab:min-wage}. The ATT is $-0.0490$, with an asymptotic
standard error of $0.0171$ ($P = 0.004$), suggesting a statistically
very significant reduction in teen employment. In contrast, the
cluster-jackknife standard error is $0.1960$ ($P = 0.8045$).

\begin{table}[t]
\centering
\caption{Effect of minimum wage on teen employment}
\label{tab:min-wage}
\vskip -6pt
\begin{tabular*}{\textwidth}{@{\extracolsep{\fill}} l d{2.4} d{2.4} d{2.4}
d{2.4} d{2.4} }
\toprule
Method & \multicolumn{1}{c}{ATT} & \multicolumn{1}{c}{Std.\ error} 
& \multicolumn{1}{c}{$P$ value} & \multicolumn{1}{c}{CI lower} 
& \multicolumn{1}{c}{CI upper} \\
\midrule
\multicolumn{6}{l}{Panel A: Original estimation sample} \\
CSDID asymptotic & -0.0490 & 0.0171 & 0.0040 & -0.0825 & -0.0154 \\
Cluster jackknife (CV$_{3}$) & -0.0490 & 0.1960 & 0.8045 & -0.4504 & 0.3525 \\
\midrule
\multicolumn{6}{l}{Panel B: Excluding Indiana} \\
CSDID asymptotic & 0.1476 & 0.1385 & 0.2860 & -0.1238 & 0.4190 \\
Cluster jackknife (CV$_{3}$) & 0.1476 & 0.5159 & 0.7770 & -0.9108 & 1.2060 \\
\bottomrule
\end{tabular*}
\end{table}

\begin{figure}[t]
	\centering
	\includegraphics[width=0.8\textwidth]{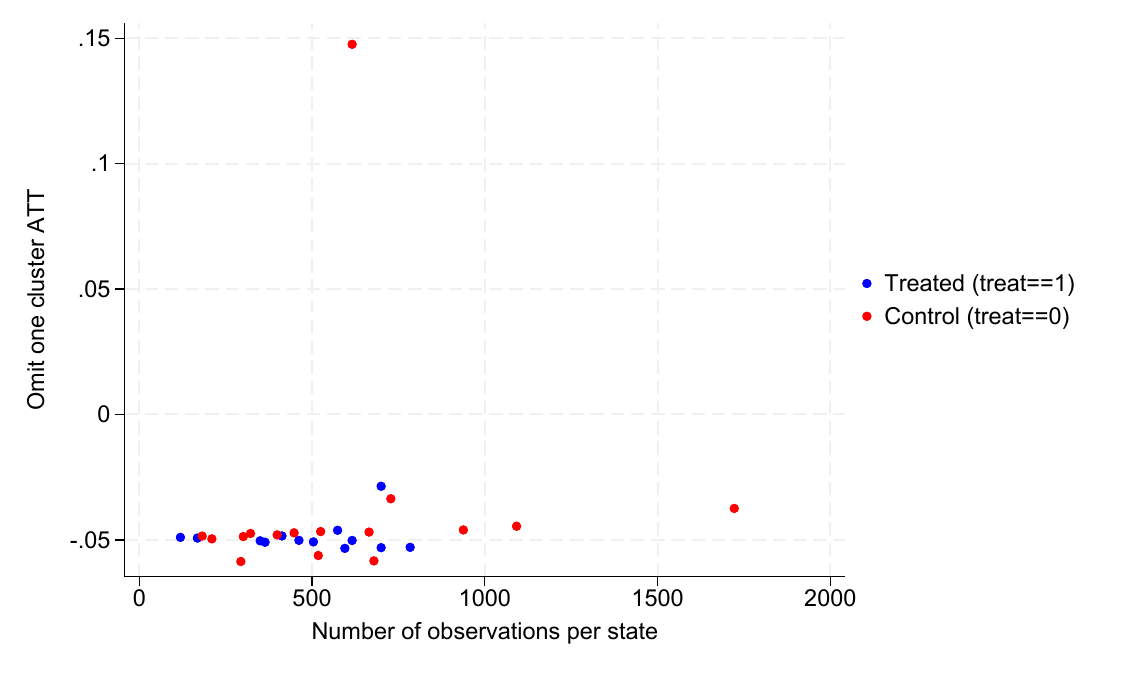}
	\caption{ATT$^{(h)}$ vs.\ cluster size for minimum wage example}
	\label{fig:att-min}
\end{figure}

Given the striking difference between the asymptotic and jackknife 
standard errors, we investigate the influence of each state on the 
aggregate effect. In \Cref{fig:att-min}, we plot these estimates
against the number of observations in the state/cluster. The red dots
indicate control states, and the blue dots indicate treated states. It
is evident from this figure that one control state is an extreme
outlier. The outlier state is ID~6 (Indiana). Although Indiana has an
intermediate number of observations, dropping it results in a
subsample ATT estimate of~$0.1476$, whereas all other leave-one-out
ATT estimates are negative and close in magnitude to the full sample
estimate.

The extreme influence of Indiana suggests that it is a high-leverage 
outlier. We therefore drop Indiana and repeat the analysis with the 
remaining 28 states. These results are presented in Panel~B of 
\Cref{tab:min-wage}. With Indiana excluded, the ATT is~$0.1476$. The
asymptotic standard error increases to 0.1385, and the jackknife 
standard error increases to~$0.5159$; in both cases, the coefficient
is no longer statistically significant. Evidently, as is the case when
using the jackknife for cluster-robust inference in least squares 
regression, it is highly instructive to inspect the full vector of 
omit-one-cluster estimates to ensure that results are not driven by a
single cluster. Our \texttt{summclust} package \citep*{MNW-influence}
provides these estimates for TWFE and other least squares regression
models. Our \texttt{csdidjack} package (see below) makes them
available for CSDID estimation of the~ATT.

\section{Software Packages}
\label{sec:software}

We provide open-source software implementations of the cluster
jackknife (CV$_3$) inference procedure for both \texttt{Stata} and 
\texttt{R} users.

\subsection{\texttt{Stata}: \texttt{csdidjack}}

The \texttt{csdidjack} package is a post-estimation command for use
with the \texttt{csdid} command \citep*{rios2021csdid} and with the
built-in \texttt{Stata} command \texttt{hdidregress}. The package
supports the \texttt{agg(simple)}, \texttt{agg(group)}, and
\texttt{agg(calendar)} aggregation schemes.

To install or update \texttt{csdidjack} in \texttt{Stata}, run:
\begin{verbatim}
net install csdidjack, \\\
from("https://raw.githubusercontent.com/liu-yunhan/csdidjack/main/") replace
\end{verbatim}
Once a model has been estimated using \texttt{csdid} or
\texttt{hdidregress} and \texttt{estat aggregation}, users can run:
\begin{verbatim}
csdidjack
\end{verbatim}
This command returns CV$_3$ standard errors, $t$-statistics, $P$~values,
and confidence intervals. For further documentation, type 
\texttt{help csdidjack} in \texttt{Stata}. Source code is available at
\url{https://github.com/liu-yunhan/csdidjack}.

\subsection{\texttt{R}: \texttt{didjack}}

The \texttt{didjack} package provides CV$_3$ jackknife inference
for users of the \texttt{did} package by Callaway and Sant’Anna 
in~\texttt{R}. 

To install the package from GitHub, run:
\begin{verbatim}
remotes::install_github("liu-yunhan/didjack")
\end{verbatim}
An example of its usage is:
\begin{verbatim}
library(did)
library(didjack)
gt  <- att_gt(yname = "log_wage_dm", tname = "year", gname = "first_treat",
data = df, clustervars = "state", panel = FALSE,
est_method = "reg", bstrap = FALSE)
ag  <- aggte(gt)
res <- didjack(ag)
summary(res)
\end{verbatim}
Full documentation and source code are available at
\url{https://github.com/liu-yunhan/didjack}.

\section{Conclusions}
\label{sec:conclusions}

Obtaining reliable cluster-robust inferences for DiD estimates of 
treatment effects has been the subject of much research for over two
decades. A great deal of progress has been made, but there are still
cases (such as few clusters and/or few treated clusters) where
inference can be quite unreliable; see \citet*{MNW-guide}.

In recent years, well-founded concerns over staggered adoption have
led to new ways to estimate treatment effects using the so-called 
``modern DiD'' methods. Unfortunately, for these new estimators, the
concern over obtaining unbiased estimates of the ATT has shifted focus
away from the question of statistical inference. This short paper
highlights the fact that the long-standing problems of few clusters
and few treated clusters for TWFE estimation can be severe with the
CSDID estimator of \citet{callaway2021difference}. We propose to use
cluster-jackknife standard errors instead of the ones suggested in
that paper. Simulations strongly suggest that this can lead to much
more reliable inferences.

\bibliography{mwsshrc}
\addcontentsline{toc}{section}{\refname}

\end{document}